\documentclass[journal=jacsat,manuscript=article]{achemso}
\usepackage{chemformula} 
\usepackage[T1]{fontenc} 
\usepackage{orcidlink}
\usepackage{multirow, tabularx, booktabs}
\usepackage{color}
\usepackage{mathrsfs}
\usepackage{amsmath}
\usepackage{hyperref}

\author{Xinfeng Chen\orcidlink{0000-0003-2579-273X}}
\affiliation{Frontier Institute of Science and Technology, State Key Laboratory of Electrical Insulation and Power Equipment, Xi’an Jiaotong University, Xi’an 710049, China.}

\author{Ning Ding}
\affiliation{School of Physics, Southeast University, Nanjing 211189, China.}

\author{Paolo Barone}
\affiliation{CNR-SPIN, Area della Ricerca di Tor Vergata, Via del Fosso del Cavaliere 100, I-00133 Rome, Italy.}

\author{Carlo Rizza}
\affiliation{Department of Physical and Chemical Sciences, University of L'Aquila, Via Vetoio I-67100 Coppito, L'Aquila, Italy.}

\author{Shuai Dong}
\affiliation{School of Physics, Southeast University, Nanjing 211189, China.}

\author{Wei Ren}
\affiliation{Physics Department, Shanghai Key Laboratory of High Temperature Superconductors, State Key Laboratory of Advanced Special Steel, International Centre of Quantum and Molecular Structures, Shanghai University, Shanghai 200444, China.}

\author{Paolo G. Radaelli}
\email{paolo.radaelli@physics.ox.ac.uk}
\affiliation{Clarendon Laboratory, Department of Physics, University of Oxford, Oxford OX1 3PU, United Kingdom.}

\author{Gaoyang Gou\orcidlink{0000-0003-1485-8115}}
\email{gougaoyang@mail.xjtu.edu.cn}
\affiliation{Frontier Institute of Science and Technology, State Key Laboratory of Electrical Insulation and Power Equipment, Xi’an Jiaotong University, Xi’an 710049, China.}

\author{Alessandro Stroppa}
\email{alessandro.stroppa@spin.cnr.it}
\affiliation{CNR-SPIN, c/o Department of Physical and Chemical Sciences, University of L'Aquila, Via Vetoio I-67100 Coppito, L'Aquila, Italy.}

\title{Unconventional Magnetism, Sliding Ferroelectricity, and Magneto-Optical Kerr Effect in Multiferroic Bilayers}

\begin{document}

\newpage

\begin{abstract}

Antiferromagnetic (AFM) materials provide a platform to couple altermagnetic (AM) spin-splitting with the magneto-optical Kerr effect (MOKE), offering potential for next-generation quantum technologies. In this work, first-principles calculations, symmetry analysis, and \textit{\textbf{k$\cdot$p}} modeling are employed to show that interlayer sliding in AFM multiferroic bilayers enables control of electronic, magnetic, and magneto-optical properties. This study reveals an intriguing dimension-driven AM crossover: the 2D paraelectric (PE) bilayer exhibits spin-degenerate bands protected by the [$\mathcal{C}_2 \| \mathcal{M}_c$] spin-space symmetry, whereas the 3D counterpart manifests AM spin-splitting along $k_{z} \neq 0$ paths. Furthermore, interlayer sliding breaks this $\mathcal{M}_c$ symmetry and stabilizes a ferroelectric (FE) state with compensated ferrimagnetism, where the Zeeman-like field is responsible for the nonrelativistic spin-splitting. In the FE phase, spin-orbit coupling (SOC) lifts accidental degeneracies and produces ``alternating'' spin-polarized bands through the interplay of Zeeman and Rashba effects. Crucially, spin polarization, ferrovalley polarization ($\Delta$$E_{V}$), and the Kerr angle ($\theta_{k}$) can all be reversed by switching either sliding ferroelectricity or the N\'{e}el vector. Our findings reveal the rich coupling among electronic, magnetic, and optical orders in sliding multiferroics, illustrating new prospects for ultralow-power spintronic and optoelectronic devices.

\end{abstract}


\section{INTRODUCTION}

Altermagnets (AMs) are compensated magnets in which similar to ferromagnetic (FM) materials, Kramers spin-degeneracy is lifted in such a way as to allow $\vec{k}$/$-\vec{k}$ symmetric spin-split energy bands in certain regions of the Brillouin zone\cite{001163408400003,001312751500001,000902137200001}. AMs are a subset of compensated magnets (historically called antiferromagnets, AFMs) characterized by two symmetry-related collinear but opposite spin-sublattices. In order to allow non-relativistic spin-splitting, the two spin-sublattices must not be related by either a translation ($\vec{t}$, leading to time-reversal $\mathcal{T}$ symmetry) or inversion ($\mathcal{I}$, leading to $\mathcal{TI}$ symmetry). Instead, opposite spin-sublattices in AMs are connected by real-space rotation ($\mathcal{R}$) or mirror ($\mathcal{M}$) operations\cite{000902137200001,001140896900001}. Unlike the case of $\mathcal{T}$ and $\mathcal{TI}$ symmetries, this relation does not always lead to exact symmetries of the magnetically ordered phase. The exact or approximate [$\mathcal{C}_2 \| \mathcal{R}$] (in spin group framework\cite{000608669800007,001445038600001}) or $\mathcal{TR}$ (in Shubnikov framework\cite{Paolo}) symmetries are hallmarks of AMs and offer a basis for their classification. Since their recognition as a distinct class, AMs have been intensively studied to confirm their distinctive electronic structures. Initial reports focused on RuO$_{2}$\cite{001262232800001} (with later debate about its magnetism\cite{6666666666}), CrSb\cite{000943379700001} and Mn$_{5}$Si$_{3}$\cite{001245213500034}. Moreover, several reports made a connection between AMs and non-collinear magnets such as MnTe$_{2}$, which were known to support spin-splitting that are very similar to those of AMs\cite{001163408400003}. By combining FM-like and AFM-like magnetic properties in a single phase, AMs exhibit unique physical characteristics and offer application advantages beyond those of other magnetic materials\cite{000865310200001,25943312}.

An important strand of this research focused on the role of SOC. In the absence of SOC, AMs support spin-splitting of non-relativistic origin that can be as large as 1 eV, the resulting band structure being insensitive to the direction of the N\'{e}el vector. When SOC is introduced, additional band structure modifications and momentum-space splitting take place, which depend on the direction of the magnetic moment and are generally associated with the emergence of a small degree of spin non-collinearity. This phenomenon was observed in MnTe\cite{001163408400002} and is sometimes called `weak' altermagnetism, in analogy to the well-known effect leading to `weak ferromagnetism'. When SOC-dependent weak effects are taken into account, they may lead to further classifications of AMs, such as the one recently proposed by S.-W. Cheong\cite{001147635900001}. In general, spin-splitting associated with both `strong' and `weak' altermagnetism can lead to unique properties, such as spin current and spin torque generation\cite{000754669000001,000807542100004}, the crystal hall effect\cite{000879720400003}, giant tunneling magnetoresistance\cite{000754669000001,001286218000004} and chiral magnons\cite{001153468500007,0003903014000836545}, thus attracting great attention in the fields of spintronics, magnetic storage, dissipationless magneto-electrics and quantum computing\cite{25614631,001302265700002,001445038600001}.

The magneto-optical Kerr effect (MOKE) is the rotation of linearly polarized light upon reflection from magnetic materials; the rotation angle is termed Kerr angle, $\theta_{k}$. This effect provides a way to probe magnetic ordering and investigate key magnetic properties\cite{000411659000003}. Initially, MOKE was observed exclusively in FM and ferrimagnetic (FiM) materials with non-zero magnetization\cite{A1994MW35100033}. More recently, MOKE was observed in nearly compensated non-collinear AFMs\cite{000363512400005}, and also in collinear AM candidates such as RuO$_{2}$ and CoNb$_{3}$S$_{6}$\cite{000669001000002}, where $\theta_{k}$ reverses with crystal handedness. Time-resolved MOKE measurements have been further demonstrated to be effective for ultrafast tracking of spin dynamics in epitaxial MnTe films\cite{001312751500001}. Combined with angle-resolved photoemission spectroscopy (ARPES) \cite{001181900200001}, MOKE provides a versatile and experimentally accessible probe for unambiguous detection of AM.

For a non-zero Kerr angle to be allowed, a necessary condition is that $\mathcal{T}$ and $\mathcal{TI}$ symmetries be simultaneously broken\cite{000885000400002,001105561900001}, as is the case for AMs. However, a second necessary condition is that FM ordering should be allowed. Therefore, only a subset of AMs is MOKE-active\cite{000500914600003,001147635900001}. Moreover, the MOKE arises only via SOC, because FM moments (even weak ones) are forbidden by AM spin group symmetries. Consequently, FE polarization coupled to spins through SOC can lead to additional multifunctional responses\cite{25614631}. In some cases, FE polarization can lead to a transition from non-AM and MOKE-inactive to MOKE-active AM\cite{001463466500001,001494648600001,001459669800001}. For example, the centrosymmetric MnPSe$_{3}$ monolayer displays collinear AFM magnetic ordering, where spin-degeneracy is protected by $\mathcal{TI}$ symmetry. By introducing out-of-plane polarization, via either an applied vertical electric field or Janus structure formation (substituting Se with S in one atomic layer)\cite{MnPSe}, the $\mathcal{TI}$ symmetry is broken, and a mirror symmetry relates the magnetic moments of two oppositely oriented Mn spin-sublattices. This transformation converts the centrosymmetric AFM monolayer into a 2D polar AM material, while simultaneously activating the MOKE. Another promising strategy for breaking $\mathcal{I}$ symmetry in 2D materials involves stacking two monolayers in specific interlayer configurations, leading to net out-of-plane ferroelectricity\cite{000453488800031,001180085100001,001181249100001,001441248700001}. Consequently, FE polarization and magnetic ordering provide additional degrees of freedom for the simultaneous tuning of the electronic structure and the MOKE response.

In this paper, we perform first-principles calculations and symmetry analysis to investigate the interplay between magnetism, sliding ferroelectricity, and the MOKE in the 2D AFM H$^{'}$-Co$_{2}$CF$_{2}$ multiferroic. In the PE phase, the bilayer exhibits spin-degenerate bands without SOC, protected by the mirror symmetry $\mathcal{M}_c$\cite{25614631}. In contrast, its bulk phase with 3D periodicity displays AM spin-splitting along reciprocal space paths with $k_{z} \neq 0$. This highlights a dimension-driven AM crossover between 3D bulk and 2D layered forms. In the FE bilayer with out-of-plane sliding ferroelectricity, the emergence of compensated ferrimagnetism lifts spin-degeneracy even in the absence of SOC. Upon inclusion of SOC, the momentum-dependent spin-splitting gives rise to spin polarization changing within each energy band across the Brillouin zone. We refer to this peculiar behavior as ``alternating'' spin-polarized bands, which arises from the interplay between Rashba and Zeeman effects. In addition, we observe pronounced spin-valley locking with nontrivial spin-texture. Remarkably, the sign of both the ``alternating'' spin-polarized bands and $\theta_{k}$ can be reversed either by switching the FE polarization or by reversing the N\'{e}el vector. These findings demonstrate how the FE and AFM orderings provide versatile degrees of freedom for engineering MOKE in 2D materials.

\section{RESULTS AND DISCUSSION}

\subsection{Crystal structure and magnetic configuration}

Co$_{2}$CF$_{2}$ belongs to the MXene material family, where each monolayer consists of five closely stacked atomic layers: F-Co-C-Co-F. In its stable H$^{'}$ phase, the H$^{'}$-Co$_{2}$CF$_{2}$ monolayer exhibits out-of-plane FM ordering and polar displacements of the Co cations relative to the anionic plane (see \textbf{Figure S1}), making it a multiferroic material with both ferromagnetism and out-of-plane ferroelectricity\cite{000829495100001}. MXene can be readily prepared as multilayers. Depending on the interlayer stacking sequence, the structural symmetry and material properties of 2D multilayers can differ significantly from their monolayer counterparts (e.g., 2H-MoS$_{2}$)\cite{000332059200004}.

\begin{figure}[h]
\centering
\includegraphics[width=0.75\linewidth]{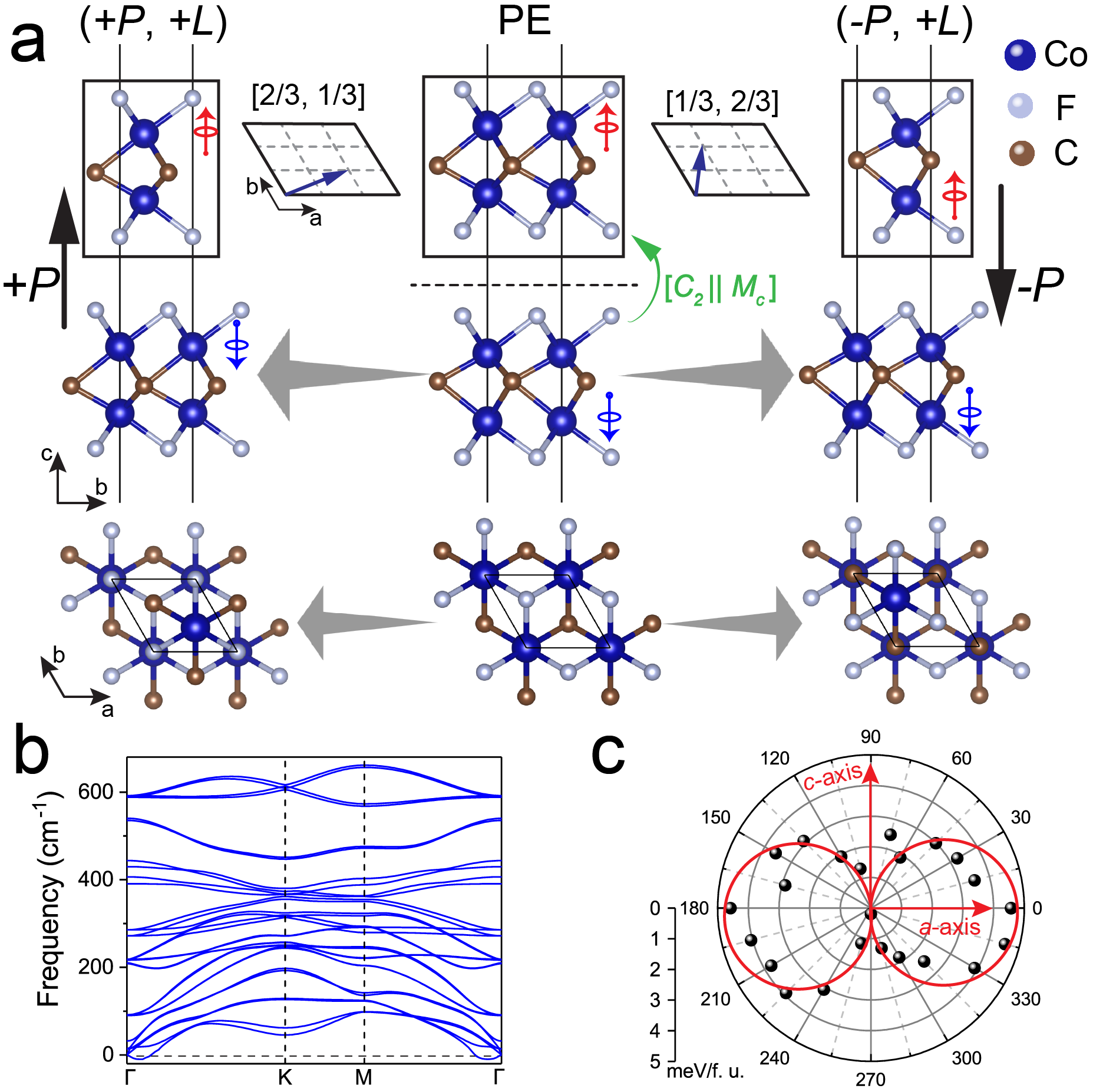}\vspace{-3pt}
\caption{(a) Schematic diagrams of the transition from the PE to FE phase ($\pm$$P$ states) by the [2/3, 1/3] and [1/3, 2/3] interlayer sliding within H$^{'}$-Co$_{2}$CF$_{2}$ bilayer. Both side and top views are shown. The unit cell of H$^{'}$-Co$_{2}$CF$_{2}$ monolayer is outlined in black, with Co, C, and F atoms colored in blue, brown, and silver, respectively. Red and blue circled arrows represent the direction of magnetic moments for spin-active Co cations. In the FE phase, out-of-plane polarization direction $P$ is shown by black arrows. (b) Simulated phonon spectrum for the ($+P$, $+L$) state of the FE bilayer, indicating no imaginary frequencies and structural instability. (c) Calculated magnetic anisotropy energy (MAE) for the ($+P$, $+L$) state, showing the magnetic easy axis along the out-of-plane $c$-axis after rotating the N\'{e}el vector within the $ac$ plane.}
\label{Figure 1}
\end{figure}

The study begins by establishing the reference PE phase for the H$^{'}$-Co$_{2}$CF$_{2}$ bilayer. As shown in \autoref{Figure 1} (a), the PE phase consists of two monolayers with an $AA^{'}$ stacking sequence and antiferroelectric (AFE) configuration. Due to the presence of a mirror plane $\mathcal{M}_c$ between the two monolayers, the corresponding space group is the acentric non-polar $P\bar{6}m2$. The lowest-energy configuration consists of two antiparallel FM monolayers, yielding a perfectly compensated AFM ordering with type-III magnetic space group (MSG) $P\bar{6}^{'}m^{'}2$ (detailed results are tabulated in \textbf{Table S1}). It is important to note that only the Co cations farther from the C atomic layer exhibit non-zero magnetic moments. Here, they are referred to as spin-active Co, labeled with circled arrows in \autoref{Figure 1} (a). In contrast, the Co cations in the other layer have zero magnetic moments. This coexistence of spin-active and spin-inactive Co cations is the result of differences in orbital occupancy, induced by the varying lengths of Co-C bonds\cite{000829495100001}. In the PE phase, two spin-sublattices with opposite Co magnetic moments are connected by the [$\mathcal{C}_2 \| \mathcal{M}_c$] spin-space operation.

However, the PE phase of the bilayer is unstable, as it undergoes a spontaneous PE--FE phase transition induced by the relative sliding of the adjacent layers. As illustrated in \autoref{Figure 1} (a), after sliding the top monolayer toward the [2/3, 1/3] or [1/3, 2/3] in-plane positions of the bottom monolayer, two stable and equivalent FE structures with $AB$ stacking sequence are formed, corresponding to the acentric polar $P3m1$ space group resulting from the breaking of $\mathcal{M}_c$ reflection. As a result, the Co cations of one monolayer are positioned exactly below or above the F anions of the other monolayer, leading to net interlayer dipoles. Note that the C anions in both monolayers are displaced along the same out-of-plane direction. The estimated polarization is $P$ = $\pm$2.17 $\mu$C/cm$^{2}$ (see \textbf{Figure S2} for computational details\cite{000168937200100}). As shown in \autoref{Figure 1} (b), the dynamical stability of the FE bilayer is confirmed by the absence of imaginary frequencies in the phonon band structure.

The lowest-energy configuration is an antiparallel AFM arrangement of the FM monolayers (see \textbf{Table S1}). To determine the magnetic easy axis of the FE-AFM bilayer, we evaluate the magnetic anisotropy energy (MAE) by rotating the N\'{e}el vector within the $ac$ crystallographic plane. \autoref{Figure 1} (c) shows an out-of-plane easy axis for the FE-AFM phase. Accordingly, the MSG is $P3m^{'}1$ (\textbf{Figure S3}), which also permits ferromagnetism. As the FE-AFM bilayer maintains its insulating character, the net magnetization remains zero by virtue of the Luttinger theorem\cite{luttingerI,luttingerII}, thus realizing a fully compensated ferrimagnet not enforced by symmetry\cite{mazin2022}. The resulting multiferroic phase is characterized by the out-of-plane FE polarization $P$ and N\'{e}el vector $\textbf{L}= \textbf{M}_{1} - \textbf{M}_{2}$, where the subscripts 1 and 2 correspond to upper and lower monolayers respectively. Within a Landau-theory framework and taking the $P\bar{6}^{'}m^{'}2$ PE phase as the reference structure, four iso-energetic symmetry-equivalent states can be predicted in the H$^{'}$-Co$_{2}$CF$_{2}$ multiferroic bilayer, characterized by ($+P$, $+L$), ($+P$, $-L$), ($-P$, $+L$) and ($-P$, $-L$) (see \textbf{Figure S4}), mutually related by the broken-symmetry elements $\mathcal{T}$, $\mathcal{M}_c$ and their combination $\mathcal{T}\mathcal{M}_c$. In experiments, the two equivalent FE states can be achieved by applying an external electric field or by optical and mechanical means\cite{ashoori_science2024,zhicheng_prl2025,yurong_prl2024}, while the reversal of N\'{e}el vector can be accomplished through the current-induced spin-orbit torque method\cite{000369291600036}.


\subsection{Dimension-driven AM crossover in PE phase}

\begin{figure}[h]
\centering
\includegraphics[width=0.9\linewidth]{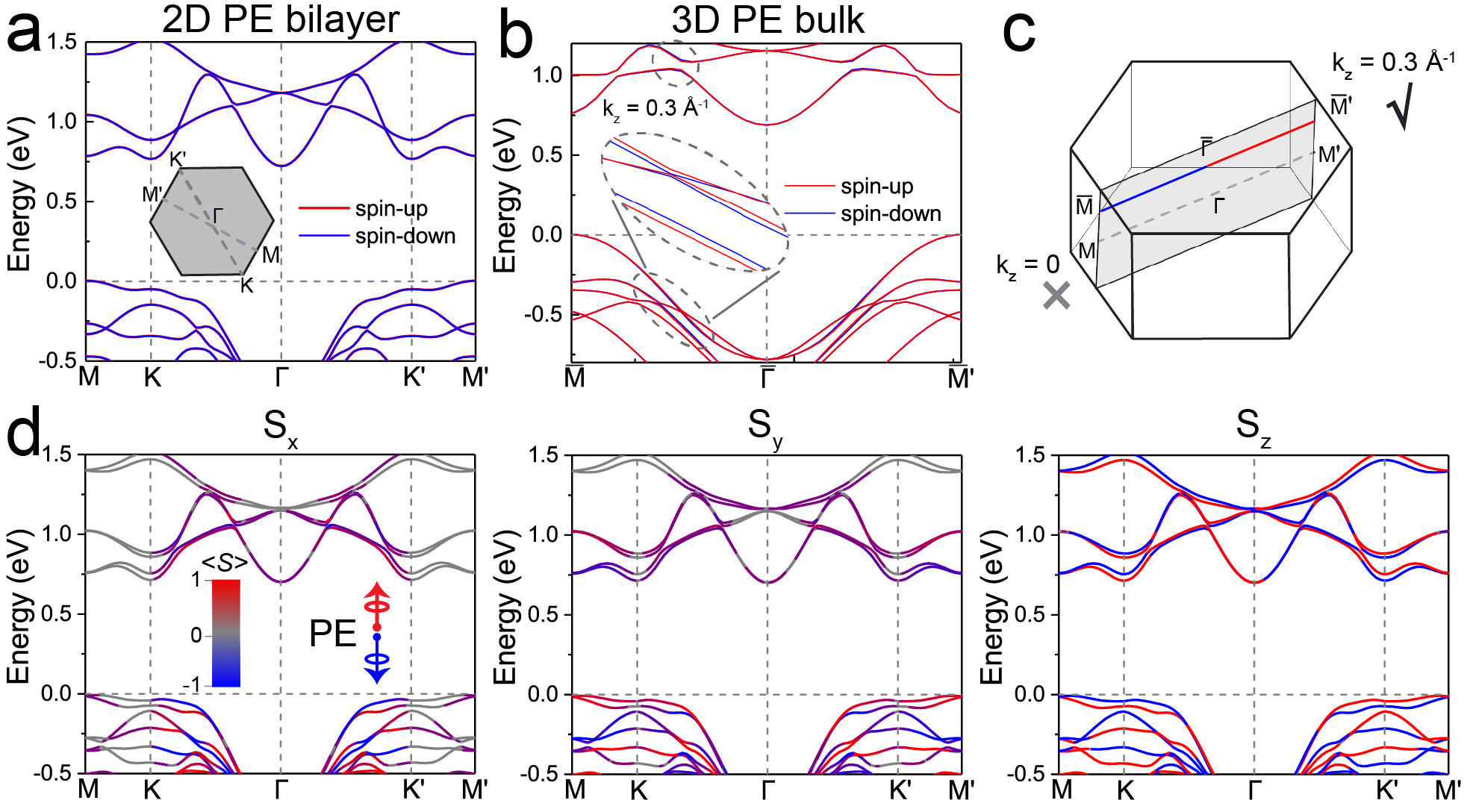}\vspace{-3pt}
\caption{(a) Calculated non-relativistic spin-resolved band structure for the 2D PE phase of the H$^{'}$-Co$_{2}$CF$_{2}$ bilayer at $k_{z} = 0$. The inset shows the high-symmetry reciprocal paths in the 2D Brillouin zone. (b) Calculated non-relativistic spin-resolved band structure for the 3D PE phase of the H$^{'}$-Co$_{2}$CF$_{2}$ bulk along the reciprocal path at $k_{z} = 0.3$ \AA$^{-1}$. (c) Schematic of the 3D Brillouin zone along the $\overline{\mathrm{M}}$--$\overline{\Gamma}$--$\overline{\mathrm{M'}}$ (M--$\Gamma$--M$^{'}$) paths at $k_{z} = 0.3$ \AA$^{-1}$ ($k_{z} = 0$), showing the corresponding spin split bands. Red and blue colors indicate opposite spin components. Spin-splitting occurs only when $k_{z} \neq 0$. (d) Relativistic spin-resolved band structure for the 2D PE bilayer with a $+L$ N\'{e}el vector. The color bar indicates the expectation values of the spin components from spin-active Co atoms. The Fermi level is set at 0 eV.}
\label{Figure 2}
\end{figure}

Spin-space group (SSG) analysis predicts that the PE bilayer, with [$\mathcal{C}_2 \| \mathcal{M}_c$] spin-space symmetry, should be AM exhibiting non-relativistic spin-splitting at $k$-points not invariant under $\mathcal{M}_c$\cite{000902137200001,000865310200001}. However, as depicted in \autoref{Figure 2} (a), the calculated non-relativistic band structure is completely spin-degenerate. This apparent contradiction can be resolved by noting that all crystalline quasi-momenta of the 2D Brillouin zone are indeed invariant under the $\mathcal{M}_c$ reflection, thus enforcing Kramers-like spin-degeneracy of the full band structure. When out‑of‑plane translation symmetry is restored, as in a 3D material possessing the same magnetic point group (MPG), the [$\mathcal{C}_2 \| \mathcal{M}_c$] symmetry forces the spin-splitting to be antisymmetric under $k_z \rightarrow -k_z$, protecting Kramers‑like degeneracy only on the $k_z = 0$ plane\cite{25614631}. Consistently, \autoref{Figure 2} (b) demonstrates that AM splitting emerges in the 3D bulk phase (\autoref{Figure 2} (c)), despite the fact that the 2D and 3D phases have the same MPG. To our knowledge, this is the first example of dimension-driven altermagnetism, with an AM-AFM crossover between 3D and 2D forms despite identical stacking. The dimension-driven altermagnetism deserves further investigations, which are beyond the purpose of the present study.

The spin-degeneracy of non-relativistic bands in the PE-AFM bilayer can be understood as follows. Each FM monolayer displays Zeeman-split bands, which are weakly modified by interactions across the van der Waals gap. The Zeeman-split band structure of the upper monolayer is mirrored in the lower monolayer as a consequence of the $\mathcal{M}_c$ symmetry, showing the same dispersion but with reversed spin polarization due to the opposite effective exchange field. Each overall spin-degenerate band thus displays a hidden Zeeman spin-splitting when projected onto mirror-partner layers, analogously to conventional $\mathcal{TI}$-symmetric AFMs with staggered magnetization on $\mathcal{I}$-related magnetic sublattices.

As shown in \autoref{Figure 2} (d), when SOC is included, the $\mathcal{M}_c$ operation is replaced by the antiunitary $\mathcal{T}\mathcal{M}_c$ operation, which enforces Kramers-like degeneracy only at the $\mathcal{T}\mathcal{M}_c$-invariant $\Gamma$ and M points, thus allowing SOC-induced spin-splitting in general. In particular, the double-degeneracy of band edges near the K(K$^{'}$) valleys is lifted, resulting in energy differences of $\Delta E_{s} = \pm 50.30$ meV and $\pm 21.25$ meV for the conduction and valence bands, respectively. These SOC-induced effects are clearly visible in the spin-resolved energy bands, where the expectation values of spins are encoded by colors (see color bars). Moreover, significant spin-splitting and unique spin-momentum locking features appear in the relativistic band structures. We recall that the non-centrosymmetric MPG of the PE phase allows both $\vec{k}$/$-\vec{k}$ symmetric (even-parity) and $\vec{k}$/$-\vec{k}$ anti-symmetric (odd-parity) bands. Due to the antiunitary $\mathcal{T}\mathcal{M}_c$ operation, the in-plane spin components ($S_{x}$ and $S_{y}$) exhibit even-parity (i.e., $E(k, S_{x, y}) = E(-k, S_{x, y})$), in agreement with the MPG analysis\cite{Paolo}. In contrast, the sign of the out-of-plane spin component $S_{z}$ has an odd parity ($E(k, S_{z}) = E(-k, -S_{z})$), and alternates along the reciprocal path of M--K--$\Gamma$--K$^{'}$--M$^{'}$. Likewise, $\mathcal{T}\mathcal{M}_c$ symmetry enforces reciprocity of the band structure; in particular, valleys K and K$^{'}$, which are precisely related by $\mathcal{T}\mathcal{M}_c$, display the same band energies but with opposite spin polarization $E$(K, $S_{z}$) = $E$(K$^{'}$, $-S_{z}$). Furthermore, by reversing the N\'{e}el vector $L$, equivalent to applying the time-reversal $\mathcal{T}$ operation, the signs of the in-plane spin components are reversed, but the $S_{z}$ component remains unchanged since $\mathcal{T}$ and $\mathcal{T}\mathcal{M}_c$ have the same effect on the odd-parity spin component (\textbf{Figure S5}).

Even though both bulk and bilayer realizations of $AA^{'}$-stacked H$^{'}$-Co$_{2}$CF$_{2}$ can be classified as AMs according to SSG symmetries, we conclude that AM behavior and non-relativistic spin-splitting can be realized only in the 3D form, whereas significant spin-splitting in the 2D case is purely driven by SOC. On the other hand, the PE-AFM phase can be classified as a type-II AM\cite{001147635900001}, due to the lack of $\mathcal{TI}$ symmetry along with the absence of FM-like behavior as enforced by the MPG $\bar{6}^{'}m^{'}2$.

\subsection{Compensated ferrimagnetism induced by interlayer sliding}

\begin{figure}[h]
\centering
\includegraphics[width=0.9\linewidth]{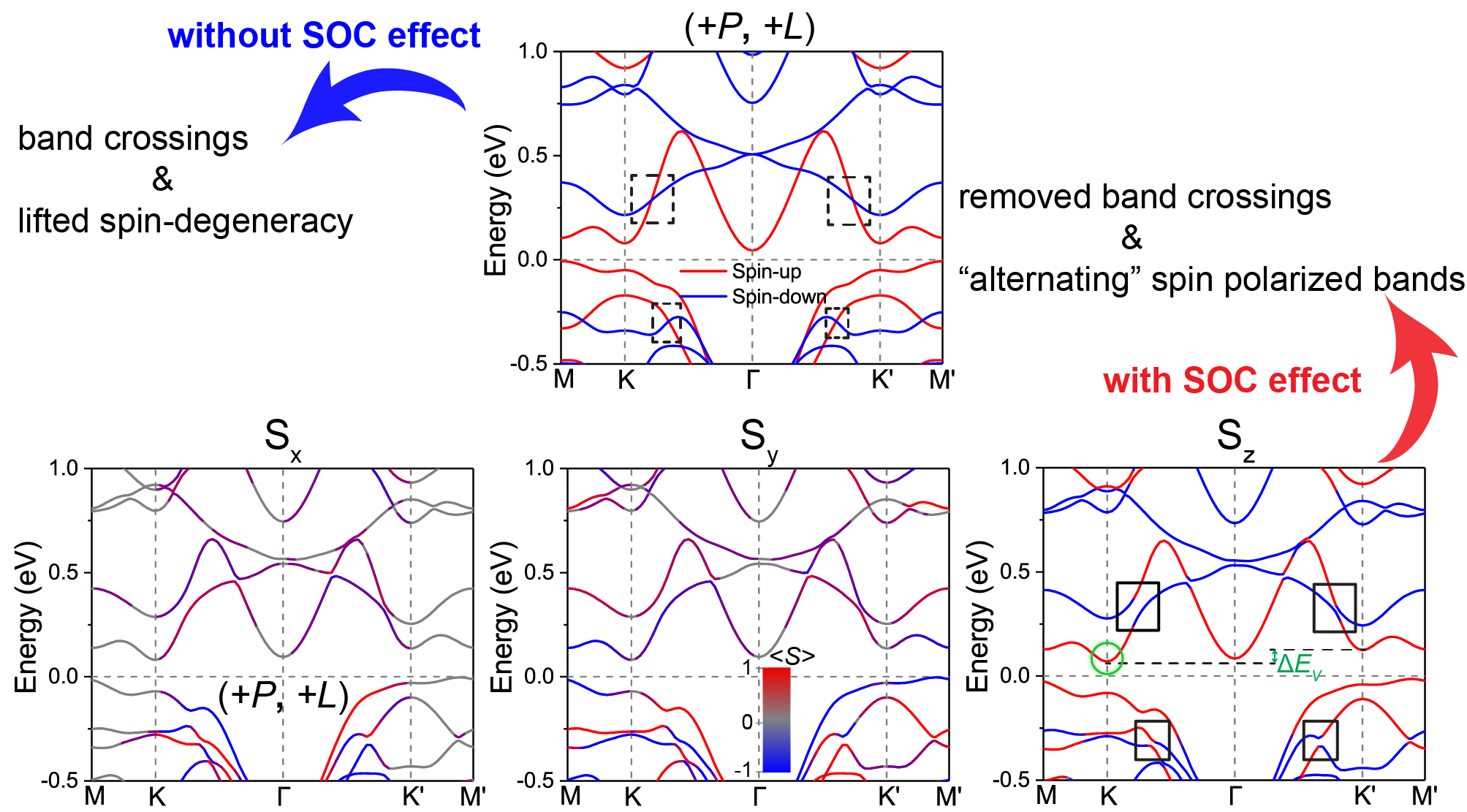}\vspace{-3pt}
\caption{Calculated spin-resolved non-relativistic (upper) and relativistic (lower) band structures for the ($+P$, $+L$) state of the FE H$^{'}$-Co$_{2}$CF$_{2}$ bilayer, respectively. Band crossings along the K--$\Gamma$--K$^{'}$ path are marked by dashed rectangles, while the removed crossings are indicated by solid rectangles. The color bar represents the expectation values of the in-plane ($S_{x}$, $S_{y}$) and out-of-plane ($S_{z}$) spin components from spin-active Co cations. Spin-valley locking at the K point is highlighted by a green circle. The Fermi level is set at 0 eV.}
\label{Figure 3}
\end{figure}

The distinct crystallographic and spin-space symmetries of the PE and FE phases result in markedly different spin-splitting behaviors in their electronic band structures. Specifically, the PE phase is characterized by the SSG of $P^{\bar{1}}\bar{6}^{1}m^{\bar{1}}2^{\infty m}1$, corresponding to the MSG of $P\bar{6}^{'}m^{'}2$, while the FE phase adopts the $P^{1}3^{1}m^{1}1^{\infty m}1$ SSG and $P3m^{'}1$ MSG\cite{001302265700002}. To elucidate the spin-splitting effects and functional properties, we focus on the ($+P$, $+L$) configuration for detailed analysis. In the absence of SOC, breaking the $\mathcal{M}_c$ symmetry by interlayer sliding removes the equivalency between the two oppositely oriented FM monolayers, resulting in distinct Zeeman-like effective exchange fields, arising from both the different local environment of FM monolayer and AFM interlayer interactions. Consequently, the energy bands arising from the two monolayers undergo largely opposite rigid energy shifts, leading to fully spin-polarized bands and sizeable Zeeman-like spin-splitting throughout the 2D Brillouin zone, including high-symmetry points $\Gamma$ and M (see upper panel of \autoref{Figure 3}). Since the PE--FE transition occurs without crossing a metallic point, the net FM magnetization is zero by virtue of the Luttinger theorem\cite{luttingerI,luttingerII}, implying that the number of spin-up and spin-down channel electrons are the same. This behavior is consistent with the classification of the system as a compensated ferrimagnet based on its SSG\cite{001302265700002,26459081}. Moreover, the MSG is compatible with FM behavior, classifying the FE-AFM bilayer as a type-I altermagnet\cite{001147635900001}. Notably, accidental crossings between spin-up and spin-down bands are observed along the K–$\Gamma$–K$^{'}$ path in reciprocal space, as highlighted by the dashed rectangles in \autoref{Figure 3} and \textbf{Figure S6}.

More intriguingly, SOC effects can further reconstruct the energy bands. As shown in the lower panel of \autoref{Figure 3} and \textbf{Figure S7}, accidental degeneracies along the K--$\Gamma$--K$^{'}$ high-symmetry path are removed, and individual bands acquire alternating spin polarization, i.e., each energy band is characterized by opposite spin polarizations alternately distributed along the $k$-path\cite{001006319400001}. We refer to this SOC-induced phenomenon as ``alternating'' spin-polarized bands (see \textbf{Figure S8} for details). Including SOC in the acentric FE-AFM bilayer induces strong band non-reciprocity, particularly pronounced at the Brillouin zone edges K and K$^{'}$, which are no longer related due to the broken $\mathcal{T}\mathcal{M}_c$ symmetry. This peculiar band reconstruction stems from two distinct SOC-mediated mechanisms, as described by a \textit{\textbf{k$\cdot$p}} model (see \textbf{Figure S9} and the detailed analysis in the \textbf{Supporting Information})\cite{000608669800007}. First, SOC introduces a valley-dependent correction to the Zeeman-like exchange field, rendering the K and K$^{'}$ valleys inequivalent (i.e., non-reciprocal) with distinct spin-splitting magnitudes. Second, the polar acentric FE-AFM bilayer enables a SOC-induced Rashba coupling. This lifts spin-degenerate band crossings (indicated by solid/dashed rectangles in \autoref{Figure 3}, \textbf{Figure S6-S7}) and mediates in-plane band spin polarization, resulting in energy bands that exhibit alternating spin components in both magnitude and direction along the momentum axis. Along the K--$\Gamma$--K$^{'}$ path, the in-plane spin components ($S_{x}$ and $S_{y}$) exhibit alternating signs. Simultaneously, the out-of-plane spin component $S_{z}$ also alternates its sign along the $\Gamma$--K and $\Gamma$--K$^{'}$ segments of the ``alternating'' spin-polarized bands. As a result, the sign of the ``alternating'' spin-polarized bands can be reversed by switching either the FE polarization $P$ or the N\'{e}el vector $L$.



\subsection{Tunable alternating spin-textures and MOKE response}

\begin{figure}[h]
\centering
\includegraphics[width=0.9\linewidth]{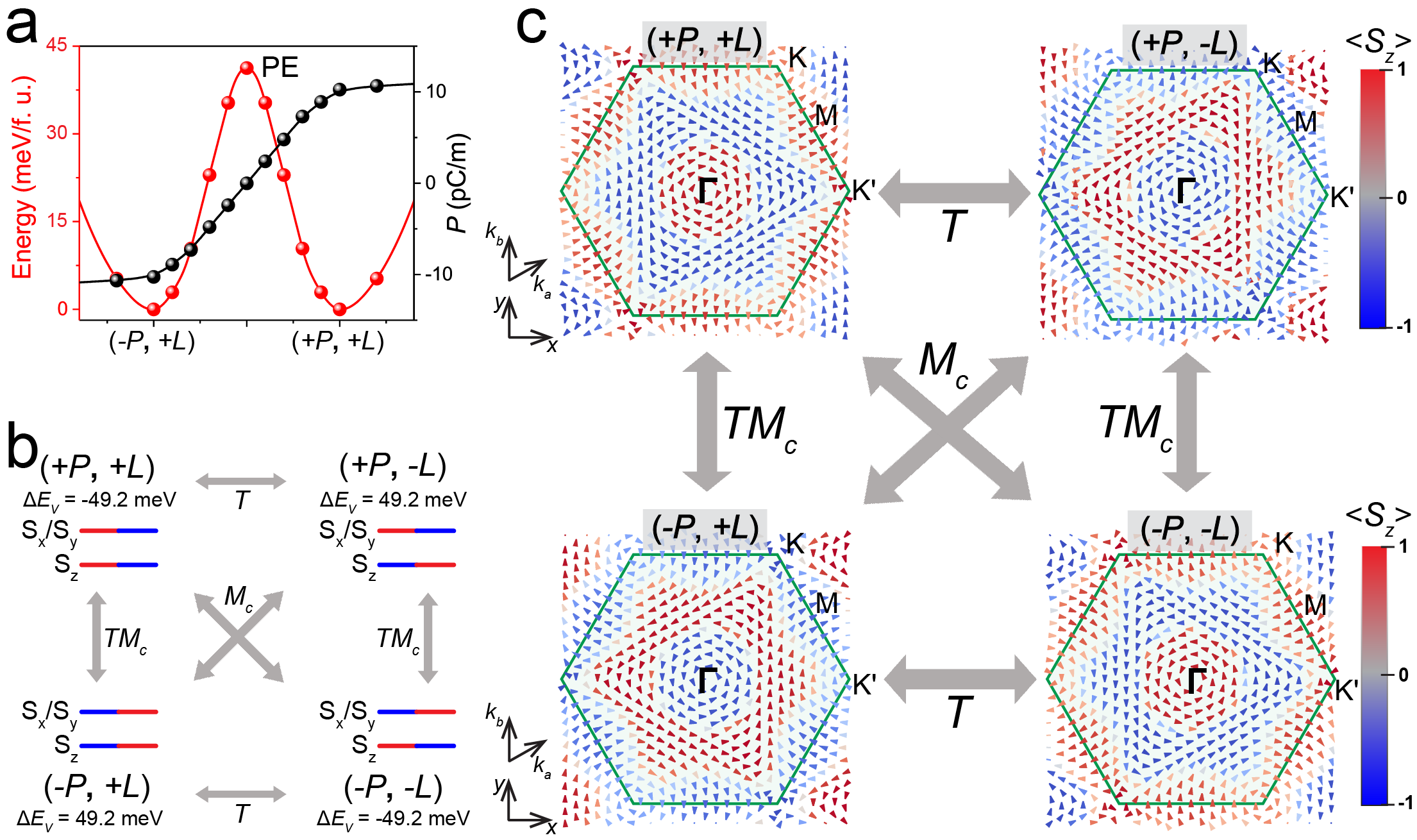}\vspace{-3pt}
\caption{(a) Energy pathway (red) and evolution of out-of-plane polarization $P$ (black) during the interlayer sliding between $\pm P$ states across the saddle point PE phase. (b) Schematic diagram illustrating the changes in the sign of spin components within the ``alternating'' spin-polarized bands and ferro-valley $\Delta E_{V}$ associated with the switching of polarization $P$ and the flipping of N\'{e}el vector $L$. The alternating red and blue bars represent the signs of the $S_{x}$, $S_{y}$, and $S_{z}$ spin components, respectively. (c) Calculated spin-textures around the conduction band edge of the four FE bilayers. In-plane spin components are depicted by arrows, and the color bars indicate the sign of out-of-plane spin component (<$S_{z}$>). The green solid lines represent the boundaries of the first Brillouin zone.}
\label{Figure 4}
\end{figure}

Since the MOKE response is governed by the ferroic state ($P$, $L$), it can be tuned concurrently through FE and magnetic means. In \autoref{Figure 4} (a), we present the energy profile and polarization evolution associated with the transition between the FE ($+P$, $+L$) and ($-P$, $+L$) states. \autoref{Figure 4} (b) illustrates the corresponding changes in the sign of the ``alternating'' spin-polarized bands marked by red and blue bars. The transition from ($+P$, $+L$) to ($-P$, $+L$) is equivalent to applying the combined $\mathcal{T}\mathcal{M}_c$ symmetry operation to the ($+P$, $+L$) state (see \textbf{Figure S3})\cite{000979037500001}. This operation simultaneously reverses both the spin-orbit and Zeeman fields, resulting in the switching of all spin components and the swapping of K and K$^{'}$ valleys. Conversely, reversing the N\'{e}el vector $L$ ($L \rightarrow -L$) is equivalent to applying a time-reversal operation $\mathcal{T}$, which inverts the Zeeman fields and exchanges the K and K$^{'}$ valleys. Consequently, only the out-of-plane spin component $S_{z}$ changes sign. Simultaneous switching of both $P$ and $L$ between ($+P$, $+L$) and ($-P$, $-L$) states corresponds to the $\mathcal{M}_c$ operation, which reverses the sign of the Rashba-like spin-orbit field. In this case, only the in-plane spin components change their signs. Interestingly, the SOC effect induces a spontaneous ferro-valley polarization, characterized by a valley splitting $\Delta E_{V} = \pm 49.20$ meV (marked by green circles in \autoref{Figure 3} and \textbf{Figure S7}). This splitting couples to the $S_{z}$ component, enabling robust spin-valley locking. Consequently, reversing the sign of $S_{z}$ via either the $\mathcal{T}$ or $\mathcal{T}\mathcal{M}_c$ symmetry operation can induce a sign reversal of $\Delta E_{V}$ at the valleys.

The modifications in spin components following either polarization switching ($P \rightarrow -P$) or N\'{e}el vector reversal ($L \rightarrow -L$) become particularly evident in the spin-texture plots. As shown in \autoref{Figure 4} (c), the spin-textures near the conduction band edges are mapped for the ($+P$, $+L$), ($+P$, $-L$), ($-P$, $+L$), and ($-P$, $-L$) states, respectively. All of these spin-texture plots exhibit three-fold rotational symmetry and Rashba-type characteristics. Consistent with the schematic diagram in \autoref{Figure 4} (b), flipping $L$ under the $\mathcal{T}$ operation preserves the orientation of in-plane spin components while the out-of-plane $S_{z}$ component reverses its sign. In contrast, switching $P$ via interlayer sliding results in the reversal of both in-plane spin components and $S_{z}$. Meanwhile, after simultaneously changing $L$ and $P$, the in-plane spin orientations reverse, while the out-of-plane spin component $S_{z}$ remains unchanged near the valley. Notably, the swapping of K and K$^{'}$ points emerges when applying $\mathcal{T}$ or $\mathcal{T}\mathcal{M}_c$ operations, which is consistent with the spin-valley locking shown in \autoref{Figure 3} and \textbf{Figure S7}. Similar results can also be observed in the spin-texture plots of their valence band edges, as shown in \textbf{Figure S10}.

\begin{figure}[h]
\centering
\includegraphics[width=1\linewidth]{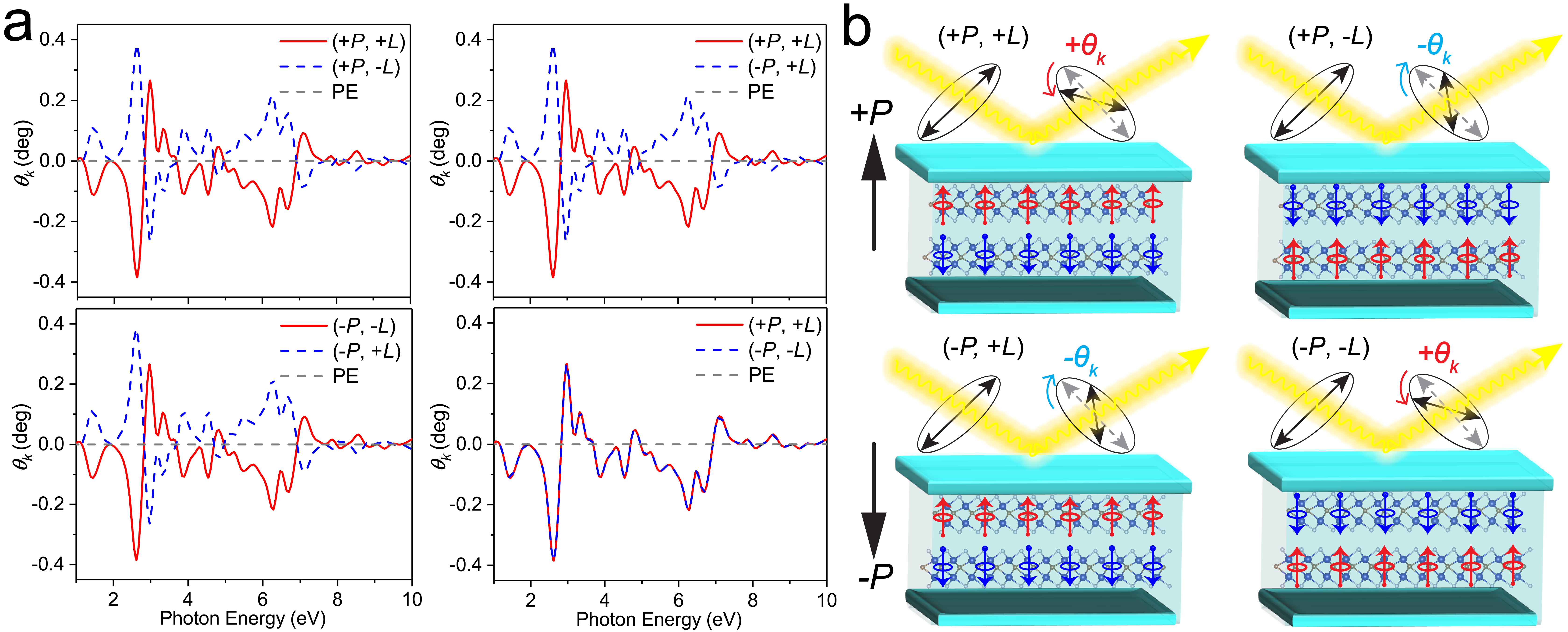}\vspace{-3pt}
\caption{(a) Simulated Kerr angle $\theta_{k}$ as a function of incident photon energy for the PE and FE phases of H$^{'}$-Co$_{2}$CF$_{2}$ bilayer. In the PE phase, MOKE is inactive with $\theta_{k} = 0$ over the entire photon energy range. In contrast, the maximal magnitude of $\theta_{k}$ occurs in the FE phase when the photon energy approaches 2.67 eV. $\theta_{k}$ reverses its sign after either switching $P$ or flipping $L$, but remains unchanged under the simultaneous change of $P$ and $L$. (b) Schematic diagram of the MOKE-based device utilizing the switching of $\theta_{k}$ in FE H$^{'}$-Co$_{2}$CF$_{2}$ bilayer. Tuning the sign of $\theta_{k}$ can be achieved by interlayer sliding between $\pm P$ states or flipping the N\'{e}el vectors between $\pm L$ states. The golden wavy lines represent incident and reflected linearly polarized light. The magnetization directions of adjacent monolayers are indicated by red and blue circled arrows. The negative and positive $\theta_{k}$ are marked in blue and red, respectively.}
\label{Figure 5}
\end{figure}

The simultaneous breaking of $\mathcal{TI}$ and $\mathcal{T}$ symmetries of the FE bilayer is responsible for the activation of MOKE. Specifically, MOKE is closely related to the frequency ($\omega$) dependent optical conductivity tensor ($\sigma$)\cite{A1992HU79900010}. For 2D layered materials lacking out-of-plane periodicity, we employ a method that combines the $\omega$-dependent effective dielectric tensor ($\varepsilon$), the effective surface optical conductivity tensor $\underline{\underline{\sigma}}^{'}$, and the effective layer thickness to simulate $\theta_{k}$ of the 2D FE H$^{'}$-Co$_{2}$CF$_{2}$ bilayer (see the \textbf{Supporting Information} for details)\cite{000979037500001}. \autoref{Figure 5} (a) shows the simulated $\theta_{k}$ for the 2D PE and FE bilayers. Without $\mathcal{T}\mathcal{M}_c$-symmetry breaking, the PE phase is MOKE-inactive with $\theta_{k} = 0$ across the spectrum. In contrast, the MOKE becomes active in the four FE states, where the maximal $\theta_{k}$ magnitudes (0.37$^{\circ}$) emerge at an incident photon energy of approximately 2.67 eV (within the visible light region). Notably, $\theta_{k}$ can be enhanced up to 14$^{\circ}$ by using a metamaterial substrate with refractive index $n_{s} = 0.78$ (\textbf{Figure S11})\cite{000500914600003}.

Since the MOKE exhibits odd-parity under time-reversal symmetry ($\mathcal{T}$-odd), the angle $\theta_{k}$ will reverse its sign when $\mathcal{T}$ operation is applied. As discussed earlier, switching the polarization $P$ or flipping the N\'{e}el vector $L$ is equivalent to applying $\mathcal{T}\mathcal{M}_c$ or $\mathcal{T}$ operations on the FE H$^{'}$-Co$_{2}$CF$_{2}$ bilayer, respectively\cite{000979037500001,000885000400002,001190821300001}. Therefore, the sign of $\theta_{k}$ will reverse under either $P$--switching [($\pm P$, $\pm L$, | $+\theta_{k}$) $\longrightarrow$ ($\mp P$, $\pm L$, | $-\theta_{k}$)] or $L$--flipping [($\pm P$, $\pm L$, | $+\theta_{k}$) $\longrightarrow$ ($\pm P$, $\mp L$, | $-\theta_{k}$)]. In contrast, when only the operation $\mathcal{M}_c$ is applied between the ($\pm P$, $\pm L$, | $+\theta_{k}$) and ($\mp P$, $\mp L$, | $+\theta_{k}$) states, the sign of $\theta_{k}$ remains unchanged. Remarkably, the transformation rules governing MOKE are identical to those governing the $S_{z}$ component in the ``alternating'' spin-polarized bands shown in \autoref{Figure 4} (b)-(c). Hence, the spontaneous and tunable $\theta_{k}$ in the FE phase follows the same behavior as the ``alternating'' spin-polarized bands, and is sensitive to the reversal of band non-reciprocity, thus providing multiple degrees of freedom to control the electronic properties of the 2D FE H$^{'}$-Co$_{2}$CF$_{2}$ bilayer\cite{001378674400003}.

Based on the tunability of $\theta_{k}$, we propose a schematic design for MOKE-active devices based on the 2D FE H$^{'}$-Co$_{2}$CF$_{2}$ bilayers. As shown in \autoref{Figure 5} (b), when linearly polarized light is incident on the ($+P$, $+L$) state, it induces a positive Kerr rotation angle (denoted as $+\theta_{k}$) in the polarization plane of the reflected light. By switching the polarization direction $P$ via an external bias or flipping the N\'{e}el vector $L$ through current-induced spin-orbit torque\cite{000369291600036}, the system transforms into the ($-P$, $+L$) or ($+P$, $-L$) state, respectively, resulting in a negative Kerr rotation ($-\theta_{k}$). This allows for electrical or magnetic control of the logic `ON' and `OFF' states. Moreover, the novel dimension-driven altermagnetism and compensated ferrimagnetism suggest that the multiferroic material H$^{'}$-Co$_{2}$CF$_{2}$ could exhibit a range of other physical properties, such as the anomalous Hall effect\cite{001543360000007}, anomalous valley Hall effect\cite{001526486000008,001530418600003,001476933000018}, Faraday rotation, and fully spin-polarized currents\cite{26459081} in its 3D PE bulk and 2D FE bilayer forms, offering a foundation for novel spintronic applications.


\section{CONCLUSIONS}

In summary, combining first-principles calculations, symmetry analysis, and \textit{\textbf{k$\cdot$p}} modeling, we systematically demonstrate the tunability of unconventional magnetism and MOKE in the H$^{'}$-Co$_{2}$CF$_{2}$ system. The 2D non-polar PE bilayer exhibits non-relativistic spin-degenerate bands due to the $\mathcal{M}_c$ mirror plane, while the non-relativistic spin-splitting associated with the [$\mathcal{C}_2 \| \mathcal{M}_c$] symmetry emerges in its 3D phase, which indicates a dimension-driven AM crossover. Breaking the $\mathcal{M}_c$ symmetry by interlayer sliding stabilizes the FE H$^{'}$-Co$_{2}$CF$_{2}$ bilayer with out-of-plane polarization and spin-split bands even without SOC. This phenomenon is attributed to compensated ferrimagnetism. When SOC is included, the FE phase characterizes ``alternating'' spin-polarized bands due to the interplay of Zeeman and Rashba effects. By tuning interlayer sliding or the N\'{e}el vector, it is possible to control the ferro-valley polarization, $S_{z}$ component, spin-texture, and $\theta_{k}$ simultaneously. Finally, we propose MOKE-active devices based on the 2D sliding FE H$^{'}$-Co$_{2}$CF$_{2}$ bilayers, enabling electromagnetic control and magneto-optical detection within the visible spectrum.

\section{COMPUTATIONAL METHODS}

First-principles calculations are performed based on density functional theory (DFT) methods as implemented in the Vienna $Ab$ $initio$ Simulation Package (VASP)\cite{1996VT67500040,1996VF38900003}. We use the Perdew–Burke–Ernzerhof (PBE) exchange-correlation functional within the generalized gradient approximation (GGA)\cite{1996VP22500044}. A plane-wave basis set within the projector augmented wave (PAW) method\cite{1994QB02200016} is employed, using a 600 eV energy cutoff. The effective Hubbard $U_{eff}$ = 2.0 eV is applied for Co-$d$ orbital\cite{000071716800040,000829495100001}. The interlayer vdW interactions are simulated using DFT-D3 method developed by Grimme\cite{000276971500005}. The crystallographic $a$, $b$ and $c$ axes refer to two in-plane and one out-of-plane (vertical) directions, respectively. 2D H$^{'}$-Co$_{2}$CF$_{2}$ bilayers are simulated as slabs containing at least 20 \AA $ $ vacuum along the $c$-axis. The Brillouin zones are sampled based on Monkhorst-Pack scheme\cite{1976BV08800009} using 12$\times$12$\times$1 $k$-point grid. The dipole correction is included in the vacuum region\cite{000080570800036}. The atomic positions and lattice parameters are fully optimized until the residual Hellmann-Feynman forces and stress are less than 0.001 eV/\AA $ $ and 0.1 kbar, respectively. The phonon spectrum is calculated based on the finite difference method\cite{000360250700001}. SOC effect is considered during the calculations. The vertical polarization in 2D bilayer can be obtained from the integration of planar-averaged screening charge density\cite{000168937200100}.

\begin{suppinfo}

The Supporting Information is available free of charge.
\begin{itemize}
  \item FE polarization of H$^{'}$-Co$_{2}$CF$_{2}$ monolayer, planar-averaged screening charge of ($+P$, $+L$), spin densities of AFM-FE phases, relative energies of FM, AFM and NM phases, AFM configurations with $-L$ N\'{e}el vector, relativistic band structure of PE phase, band structures of FE phases without SOC, band structures of FE phases with SOC, the comparison of spin-degenerate, spin-splitting and ``alternating'' spin-polarized bands, the lower conduction bands simulated by \textit{\textbf{k$\cdot$p}} model, spin-textures of valence band edges, the evolution of Kerr angle $\theta_{k}$ as a function of $n_{s}$, non-zero $\xi_{xy}$ components of FE phases, computational details of \textit{\textbf{k$\cdot$p}} model and MOKE.
\end{itemize}

\end{suppinfo}

\subsection{\textbf{Notes}}
The authors declare no competing financial interest.

\begin{acknowledgement}

The authors thank Gu Mingqiang of Sun Yat-sen University for fruitful discussions. G.Y. G acknowledges the funding support from National Natural Science Foundation of China (Grant No. 11574244). N. Ding acknowledges the support from Program for Excellent Postdoctoral Talent under Grant No. 2024ZB001, and the China Postdoctoral Science Foundation under Grant No. 2024M760423. Additional funding was provided by the European Union – NextGenerationEU under the Italian MUR National Innovation Ecosystem grant ECS00000041 - VITALITY - CUP E13C22001060006. X.F. C acknowledges the support from China Scholarship Council. This work has been funded by the European Union - NextGenerationEU, Mission 4, Component 2, under the Italian Ministry of University and Research (MUR) National Innovation Ecosystem grant ECS00000041 - VITALITY - CUP B43C22000470005. Hefei Advanced Computing Center is acknowledged for computational support.

\end{acknowledgement}


\providecommand{\latin}[1]{#1}
\makeatletter
\providecommand{\doi}
  {\begingroup\let\do\@makeother\dospecials
  \catcode`\{=1 \catcode`\}=2 \doi@aux}
\providecommand{\doi@aux}[1]{\endgroup\texttt{#1}}
\makeatother
\providecommand*\mcitethebibliography{\thebibliography}
\csname @ifundefined\endcsname{endmcitethebibliography}
  {\let\endmcitethebibliography\endthebibliography}{}

\newpage
\begin{figure}[h]
\textbf{Table of Contents}\\
\medskip
  \includegraphics[width=8cm,height=3cm]{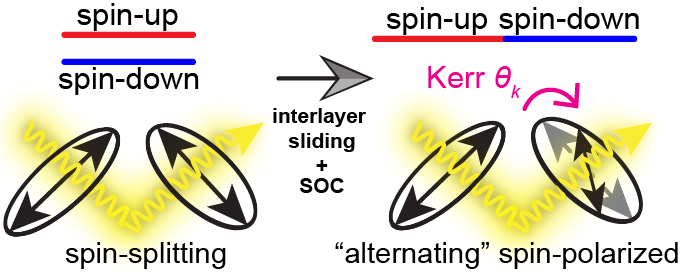}
  \medskip
  \caption*{TOC}
\end{figure}

\end{document}